\documentclass[apl, twocolumn,  amsmath,amssymb,aps]{revtex4-1}

\usepackage{graphicx}
\usepackage{dcolumn}
\usepackage{bm}
\usepackage{color}

\begin{document}

\title{III-V/Si Wafer Bonding Using Transparent, Conductive Oxide Interlayers}

\author{Adele C. Tamboli$^1$}
 \email{Adele.Tamboli@nrel.gov}
\author{Maikel F.A.M. van Hest$^1$}%
\author{Myles A. Steiner$^1$}
\author{Stephanie Essig$^1$}
\author{Emmett E. Perl$^2$}
\author{Andrew G. Norman$^1$}
\author{Nick Bosco$^1$}
\author{Paul Stradins$^1$}
\affiliation{%
 $^1$National Center for Photovoltaics, National Renewable Energy Laboratory, 15013 Denver West Pkwy, Golden, CO 80401
}%
\affiliation{
$^2$Department of Electrical and Computer Engineering, University of California, Santa Barbara, CA 93106-9560
}%

\date{\today}

\begin{abstract}
We present a method for low temperature plasma-activated direct wafer bonding of III-V materials to Si using a transparent, conductive indium zinc oxide interlayer. The transparent, conductive oxide (TCO) layer provides excellent optical transmission as well as electrical conduction, suggesting suitability for Si/III-V hybrid devices including Si-based tandem solar cells. For bonding temperatures ranging from 100$^{\circ}$C to 350$^{\circ}$C, Ohmic behavior is observed in the sample stacks, with specific contact resistivity below 1 $\Omega$\,cm$^2$ for samples bonded at 200$^{\circ}$C. Optical absorption measurements show minimal parasitic light absorption, which is limited by the III-V interlayers necessary for Ohmic contact formation to TCOs. These results are promising for Ga$_{0.5}$In$_{0.5}$P/Si tandem solar cells operating at one sun or low concentration conditions.\end{abstract}

\maketitle

Integration of III-V materials with silicon has been a persistent scientific challenge. For optoelectronic applications, it is desirable to combine the excellent optoelectronic properties of direct band gap III-V materials with technologically mature Si on which CMOS technology is based \cite{Fang2006}. For photovoltaic applications, III-V/Si tandem devices are desirable because silicon solar cells are currently the most commercially available technology, but the highest conversion efficiencies have been achieved using III-V semiconductors. Combining these two technologies offers a promising approach to scalable, efficient photovoltaic devices. Adding a wider band gap III-V top cell to a Si bottom cell (e.g., a 1.8 eV Ga$_{0.5}$In$_{0.5}$P top cell\cite{Geisz2013}) reduces the thermalization losses, boosting the cell efficiency, with a practically achievable efficiency potential of $\sim$31-32\% using known Si and Ga$_{0.5}$In$_{0.5}$P device parameters\cite{Almansouri2015}. The detailed-balance efficiency limit for a two-junction tandem cell with Si as the bottom cell is even higher, at 45\%. 
To date, the primary method used to combine Si and III-V materials has been direct growth of III-V epilayers on Si substrates \cite{Grassman2013, Yamaguchi1991}. This method is challenging due to mismatch in lattice constant and thermal expansion coefficient between Si and relevant III-V alloys as well as antiphase disorder due to polar on nonpolar epitaxy \cite{Kroemer1987}. Additionally, recent explorations have shown that exposing Si to III-V growth conditions can result in degradation of the Si bulk lifetime, compromising the efficiency potential of the Si bottom cell \cite{Garcia2014}. 

Direct wafer bonding offers a route to integration of high quality Si and III-V cells that have been separately grown and optimized. However, the realization of a two-terminal tandem device requires a bonded interface that is mechanically stable, optically transparent, and electrically conductive. Achieving all of these properties simultaneously has presented a challenge. Typical direct wafer bonding methods include either a high temperature treatment \cite{Liu2005}, where thermal expansion mismatch can be problematic, or a plasma activation step \cite{Roelkens2010}, which can result in oxidation of the semiconductors and surface damage \cite{Haussler2013, Essig2013}. Recently, equipment has been developed to enable surface activation {\em in situ} using a fast atom beam treatment, which prevents oxidation \cite{Chung1997, Essig2013, Derendorf2013}, a method that has resulted in 30\% efficiency triple junction Ga$_{0.5}$In$_{0.5}$P/GaAs//Si solar cells under concentrated sunlight \cite{Essig2015}. However, this specialized tool requires a UHV chamber equipped with fast atom or ion beam sources. 

There have been several reports of using transparent, conducting oxides (TCOs) as interlayers for bonding. Eichler, et al., measured bond strength for Si/ITO//ITO/Si bonds as a function of annealing temperature and time, using a plasma activated process\cite{Eichler2010}. ITO has also been used as an interlayer for anodic bonding of Si to glass \cite{Choi1999, Yuda2013}. High temperature ($>$500$^{\circ}$C) bonding of III-V light emitting diodes to III-V substrates has been reported\cite{Liu2005}, but such temperatures are too high for III-V/Si integration. III-V/Si integration has been reported using ZnO-mediated bonding, with a highly transparent bond interface, but the resistance of the interface was very high due to unoptimized ZnO doping\cite{Huang2014}.

We have developed an {\em ex situ} plasma-activated wafer bonding process that utilizes an amorphous indium zinc oxide (IZO) film on each semiconductor surface. This IZO film acts as a transparent contact, ensuring optical and electrical transmission between the subcells. An oxygen plasma treatment enables bonding between the IZO layers at low temperature \cite{Liang2010} without introducing damage in the semiconductor layers. While we have also used more conventional TCO materials, such as crystalline indium tin oxide, for this bonding process, we focus here on samples bonded using IZO interlayers. IZO was chosen for its excellent optical and electrical properties, as well as its smoothness\cite{Taylor2008, Taylor2005}. This TCO material forms as an amorphous film when deposited at room temperature and with compositions ranging from 55-84\% In. In this same range, the index of refraction is $\sim$2, and the conductivity can be as high as 3000 $\Omega^{-1}$cm$^{-1}$. For bonding, another key parameter is roughness; IZO films in the literature\cite{Taylor2008} have reported root-mean-squared (RMS) values $<$ 0.4 nm for a wide range of stoichiometries, which is sufficient for bonding and lower than most crystalline TCO materials. Optical absorption within the visible spectral range is insignificant at the thicknesses we use here ($<$ 20 nm). Importantly, this TCO does not change significantly upon annealing up to $\sim$600$^{\circ}$C\cite{Taylor2008}, and thus, bonding conditions (typically 100-400$^{\circ}$C) can be optimized without considering the impact on the TCO interlayer.
Here, we describe this wafer-bonding process, the resulting optical and electrical properties of the wafer-bonded interface, and its potential for tandem solar cell applications.

\begin{figure}
\includegraphics[width=0.5\textwidth]{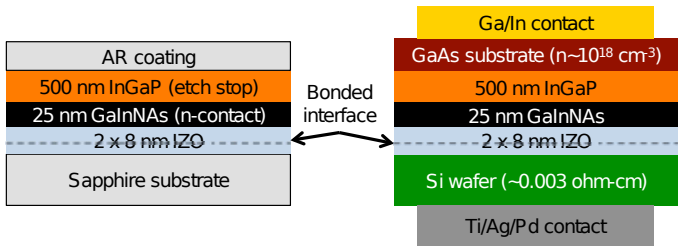}
\caption{Schematic of the test structures used here to demonstrate TCO-based wafer bonding. On the left is a device used to measure the optical properties of the bonded layers. On the right is a structure used for testing electrical properties of the bond.}
\label{schematic}
\end{figure}

Two types of sample pairs, shown schematically in Fig.~\ref{schematic}, were bonded: sapphire/IZO to GaAs/Ga$_{0.5}$In$_{0.5}$P/GaInNAs/IZO for testing optical properties, and Si/IZO to GaAs/Ga$_{0.5}$In$_{0.5}$P/GaInNAs/IZO for testing electrical properties.
Sapphire substrates used for optical characterization (Fig.~\ref{schematic}, left) were double side polished, 430 $\mu$m thick c-plane sapphire. Silicon substrates used for electrical characterization (Fig.~\ref{schematic}, right) were 500 $\mu$m thick, (100), heavily doped, single side polished n-type Si wafers (0.001-0.005 $\Omega$-cm). III-V samples consisted of 650 $\mu$m thick, (100), heavily doped n-GaAs substrates ($n$=1-4 $\times$ 10$^{18}$ cm$^{-3}$) with a nominally 25 nm thick GaInNAs contact layer \cite{Steiner2008} and nominally 500 nm thick Ga$_{0.5}$In$_{0.5}$P etch stop layer. The n-type GaInNAs layer is a heavily selenium-doped contact layer needed to make Ohmic contact to TCO layers. The n-type Ga$_{0.5}$In$_{0.5}$P layer acts as both an etch stop for substrate removal, and as a stand-in for an eventual Ga$_{0.5}$In$_{0.5}$P top cell on Si. All samples  were coated with an 8 nm thick film of amorphous IZO (70 at.\% In/30 at.\% Zn) by RF sputtering. The III-V semiconductor samples were cleaved into 1$\times$1.2 cm$^2$ pieces. Samples were cleaned in acetone, isopropanol, and tergitol and then loaded into an EVG 810 plasma activation tool. Samples were then treated with a 100 W O$_2$ plasma for 30 s. Several samples were set aside after plasma activation for atomic force microscope (AFM) measurements. Samples to be bonded were immediately stacked face to face upon unloading from the plasma chamber and loaded into a S{\"u}ss SB6e wafer bonder. Bonding was performed under vacuum, at temperatures ranging from 100-400$^{\circ}$C, for two hours. The applied pressure was 225 mbar for all samples.  

The GaAs substrates of the bonded sapphire/IZO//IZO/III-V samples were etched away chemically using a 1:1 mixture of H$_2$O$_2$ and NH$_4$OH, similar to the process described in ref. \cite{Li2013JAP}.
A wax coating around the sample edges prevented parasitic etching of IZO and GaInNAs layers. The resulting sample structures consisted of 500 nm Ga$_{0.5}$In$_{0.5}$P, 25 nm GaInNAs, 16 nm of IZO, and a 430 $\mu$m sapphire substrate (Fig.~\ref{schematic}, left). MgF$_2$ (75 nm)/ZnS (40 nm) antireflection coatings (ARC) were then deposited on the Ga$_{0.5}$In$_{0.5}$P surface. Optical properties of these samples were measured using a Cary 5000 UV/Vis/NIR transmission/reflection measurement tool. For testing the electrical conductivity through the Si/IZO//IZO/III-V samples, Ti/Ag/Pd metal stacks and Ga/In eutectic layers were deposited on the HF-etched Si and mechanically scribed GaAs back surfaces, respectively (Fig.~\ref{schematic}, right). Strength testing was performed on the same III-V/IZO/Si samples after etching off the metals. These samples were attached to glass plates on the Si side and Al blocks on the GaAs side using epoxy, and then a pull test was performed until the samples separated.

\begin{figure}
\includegraphics[width=0.5\textwidth]{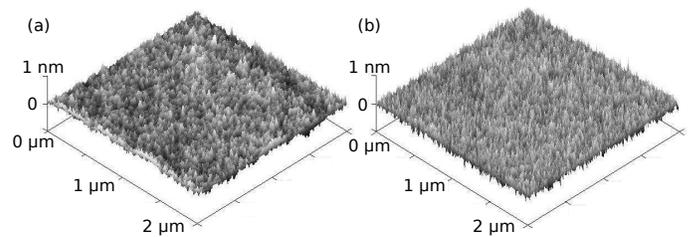}
\caption{Atomic force microscope scans of IZO/Si samples before (a) and after (b) the pre-bonding O$_2$ plasma treatment. In both cases, the RMS roughness is $<$ 0.1 nm.}
\label{AFM}
\end{figure}

\begin{figure}
\includegraphics[width=0.5\textwidth]{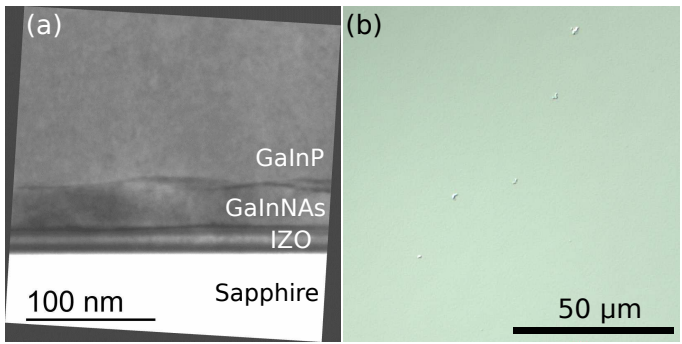}
\caption{(a) TEM imaging shows a uniform bonded interface between IZO/sapphire and IZO/III-V. The bonded interface is visible as the brighter region spanning the IZO layers. (b) Nomarski microscopy of GaInP/GaInNAs/IZO bonded to IZO/Si, after the GaAs substrate has been removed, shows few defects.}
\label{TEM}
\end{figure}

Figure~\ref{AFM} shows a comparison of AFM images that were measured on 8 nm thick IZO films on Si before and after plasma activation. Both surfaces are very flat, and AFM measurements (Fig.~\ref{AFM}) show low RMS roughness ($<$ 0.1 nm), well within the constraints required for bonding. Spectroscopic ellipsometry measurements confirm that the IZO thickness, 8 nm, did not change after plasma activation. All samples bonded at 100-400$^{\circ}$C adhered, but the Si/III-V samples bonded at high temperature cracked out of the sample plane upon cooling due to mismatched thermal expansion coefficients. 
At 400$^{\circ}$C, both the GaAs and Si cracked macroscopically through the bulk of the substrates, and some small pieces also delaminated. 
At 300-350$^{\circ}$C, small cracks formed in the 650 $\mu$m thick GaAs substrates. 
III-V/Si samples bonded at $\sim$ 200$^{\circ}$C demonstrated larger bonded areas without cracking, and exhibited an adhesion strength of 0.3 MPa. 
This bond strength is lower than optimized Si-Si plasma-activated bonds (11 MPa\cite{Li2013}) since the high temperatures necessary to convert hydrophilic to covalent bonds must to be avoided due to thermal expansion mismatch. 
The strength for III-V/Si samples bonded at 100$^{\circ}$C was lower, presumably due to large unbonded areas, which were visible as interference fringes at the III-V/sapphire samples.  
A cross sectional transmission electron microscope (TEM) image of one wafer-bonded (200$^{\circ}$C) IZO interface of a III-V/sapphire sample is shown in in Fig.~\ref{TEM}(a). The bonded interface is visible as a slightly brighter region running the length of the two dark IZO layers. Nomarski microscopy (Fig.~\ref{TEM}(b)) of a III-V/IZO//IZO/Si bonded sample after GaAs substrate removal shows few defects.

\begin{figure}
\includegraphics[width=0.4\textwidth]{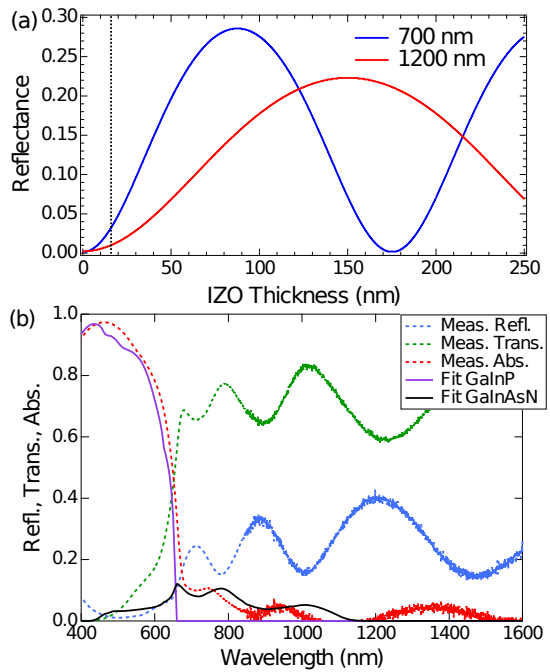}
\caption{(a) Calculated reflectance of IZO layers of varying thicknesses sandwiched between GaInP and Si. At the 16 nm total thickness used here (dotted line), we expect about 3\% reflectance at 700 nm and 1\% reflectance at 1200 nm. (b) Measured optical properties of III-V/IZO/sapphire bonded samples.  A fit, performed to the absorption data to estimate contributions from various layers, shows that majority of the absorption comes from the GaInNAs layer.}
\label{optical}
\end{figure}

The thickness of the IZO layer must be kept thin in order to mitigate reflection and absorption losses, which can be calculated using the refractive indices of the materials and thickness, $t$, of the IZO layer:

\begin{eqnarray}
R=\frac{r_1^2 + 2r_1r_2cos(\delta) + r_2^2}{1 + 2r_1r_2cos(\delta) + r_1^2r_2^2}
\end{eqnarray}

where $r_1 = (n_{GaInP}-n_{IZO})/(n_{GaInP}+n_{IZO})$, $r_2 = (n_{IZO}-n_{Si})/(n_{IZO}+n_{Si})$, and the phase shift $\delta = 4 \pi n_{IZO} t /\lambda_{vac}$\cite{Wyant2015}. The results of this calculation (Fig.~\ref{optical}(a)), show that thin layers should contribute a small amount to parasitic reflection. At 16 nm, the total thickness in this study, the predicted reflectance due to the index contrast between IZO, GaInP, and Si is 3.2\% at a wavelength of 700 nm, just below GaInP's band edge. Near Si's band edge, at 1200 nm, the predicted reflectance is 1.0\%.

A sapphire substrate was used as a stand-in for silicon in order to test the optical properties of the bonded layers, including the IZO layers, bonded interface, and GaInNAs contact layer. Transmission, reflection, and absorption ($A=1-T-R$) data for the test structures are shown in Fig.~\ref{optical}. An ARC was used to reduce the reflection from the air/Ga$_{0.5}$In$_{0.5}$P interface at the front side of the structure. Consequently, the majority of the experimentally measured reflection from the sample comes from either the III-V/IZO/sapphire interface (IZO and sapphire are nearly index matched) or the sapphire/air bottom interface. In a solar cell structure on Si rather than sapphire, these two mechanisms would not contribute significantly to reflective losses. 

At wavelengths below 660 nm, the Ga$_{0.5}$In$_{0.5}$P layer absorbs most of the light. At wavelengths greater than 800 nm, the bonded structure is relatively transparent, proving its eligibility for transmitting long-wavelength light to a Si solar cell. A fit to the data using the transfer matrix method and known optical parameters for the III-V layers shows that most of the $<$10\% parasitic absorption comes from the 25 nm thick GaInNAs contact layer, which has a 1 eV band gap. A heavily-doped contact layer is necessary to ensure Ohmic contact between the Ga$_{0.5}$In$_{0.5}$P layer and IZO contact, but could be thinned or replaced with a more optimized material. The fringes visible in the transmission and reflection data, which indicate light trapped in the Ga$_{0.5}$In$_{0.5}$P layer, suggest that further optimization of the AR coating is necessary, although this factor would be much less significant in a nearly index-matched III-V/Si structure.

\begin{figure}
\includegraphics[width=0.3\textwidth]{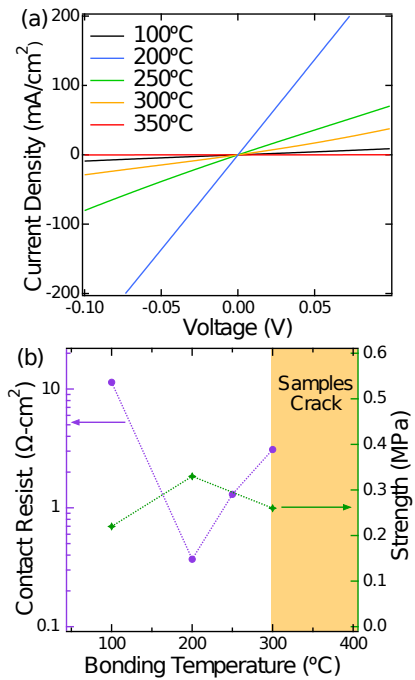} 
\caption{Electrical and strength properties of Ga$_{0.5}$In$_{0.5}$P/GaInNAs/IZO//IZO/Si stacks bonded at different temperatures. Linear IV curves are measured for all samples, but the contact resistance is minimized at $\sim$ 200$^{\circ}$C. This correlates with a maximum adhesion strength measured at 200$^{\circ}$C.}
\label{electrical}
\end{figure}

Si/IZO/III-V stacks were used to investigate the electrical properties of IZO-based bonding. Figure~\ref{electrical} shows IV curves for samples bonded at 100-350$^{\circ}$C. All samples exhibited Ohmic electrical behavior with varying resistance: at too low of a bond temperature, the bond was not sufficiently strong to yield excellent electrical properties, and at too high of a temperature, cracks inhibited current spreading. Thus, we found that the optimal condition is 200$^{\circ}$C, where $<$ 0.5 $\Omega$\,cm$^2$ was reliably achieved (5 samples measured). These results would contribute a voltage loss of $<$ 10 mV in a tandem cell operating at one sun ($\sim$20 mA/cm$^2$), sufficient for a high efficiency tandem cell. 

We report on a low-temperature bonding process using a transparent, conductive intermediate layer, which enables a transparent and conductive connection between Si and III-V semiconductors. The process works at low temperature ($\sim$200$^{\circ}$C), which is beneficial for materials with mismatched thermal expansion coefficients. The process has been demonstrated here using optically-thin, amorphous indium zinc oxide, but can also be applied to other TCOs, like indium tin oxide. This particular structure could be easily applied to tandem solar cells on silicon, and in particular, Ga$_{0.5}$In$_{0.5}$P top cells on HIT-type Si bottom cells, which already incorporate a TCO as the top surface. Such a device is promising for $>$30\% efficiency one-sun or low concentration solar cells at reasonable costs.

This work was supported by DOE EERE SETP under DE-EE00025783. Bonding was performed in the UCSB Nanofabrication Facility, a member of the NSF-funded NNIN. We thank Bobby To for providing AFM images, Waldo Olavarria for performing III-V growth, Anna Duda for ARC deposition, and Adam Stokes for TEM sample preparation.

\bibliography{refsAPL}

\end{document}